\documentclass[aps,prl,twocolumn,showpacs,groupedaddress]{revtex4}  
\usepackage{graphicx}  
\usepackage{dcolumn}   
\usepackage{bm}        
\usepackage{amssymb}   

\usepackage{multirow}   

\newcommand{\be}{\begin{equation}}
\newcommand{\ee}{\end{equation}}
\newcommand{\ba}{\begin{eqnarray}}
\newcommand{\ea}{\end{eqnarray}}
\newcommand{\ban}{\begin{eqnarray*}}
\newcommand{\ean}{\end{eqnarray*}}
\newcommand{\p}{\paragraph{}}

\newcommand{\ket}[1]{\mbox{$ | #1 \rangle $}}

\def\opone{\leavevmode\hbox{\small1\kern-3.8pt\normalsize1}}

\newcommand{\one}{\leavevmode\hbox{\small1\normalsize\kern-.33em1}}

\renewcommand{\quote}[1]{`#1'}
\newcommand{\mquote}[1]{\mbox{`#1'}}

\newcommand{\co}[1]{\hat{#1}^\dagger}

\newcommand{\fig}[1]{(Fig.~\ref{#1})}

\begin{document}

\title{Experimental Quantum Teleportation with a 3-Bell-state Analyzer}

\author{J.A.W. van Houwelingen\footnote[2]{email: jeroen.vanhouwelingen@physics.unige.ch}, A. Beveratos, N. Brunner, N.
Gisin and H. Zbinden} \affiliation{Group of Applied Physics,
University of Geneva, Switzerland\\}
\date{\today}

\begin{abstract}
We present a Bell-state analyzer for time-bin qubits allowing the
detection of three out of four Bell-states with linear optics, two
detectors and no auxiliary photons. The theoretical success rate
of this scheme is $50\%$. A teleportation experiment was performed
to demonstrate its functionality. We also present a teleportation
experiment with a fidelity larger than the cloning limit of F=5/6.
\end{abstract}
 \maketitle \p

\section{Introduction}
Bell-State Analyzers (BSA) form an essential part of quantum
communications protocols. Their uses range from quantum relays
based on teleportation \cite{Bennett1993, Bouwmeester1997,
Boschi1998, Waks2002,Jacobs2002, Marcikic2003, Collins2005} or
entanglement swapping \cite{Pan1998, Riedmatten2005} to quantum
dense coding \cite{Bennett1992,Mattle1996}. An important
restriction for BSAs is that a system based on linear optics ,
without using auxiliary photons, is limited to a 50\% overall
success rate \cite{Calsamiglia2001, Knill2001}. This important
result does not restrict the number of Bell-states that can be
measured, but only the overall efficiency of a measurement.
Nevertheless, a complete BSA is possible for at least two
different cases: the first approach uses non-linear optics
\cite{Kim2001} but this has the drawback of an exceedingly low
efficiency and is therefore not well adapted for quantum
communication protocols. Another possibility is the use of
continuous variable encoding \cite{Braunstein1998, Furusawa1998},
however, this technique has the disadvantage that postselection is
not possible. Note that postselection is a very useful technique
that allows one to use only \quote{good} measurement results and
straightaway eliminate all others  without the need for great
computational analysis.

Many experiments have been done up to date that use BSAs. In this
article a novel BSA is introduced \cite{Houwelingen2006}. It has
the maximum possible efficiency that can be obtained when using
only linear optics without ancilla photons. It is different with
respect to other BSAs since it can distinguish three out of the
four Bell-states. All of the used BSAs up to date that can reach
the maximum efficiency, without the use of ancilla photons, are
limited to two (or less) Bell-states \cite{Weinfurther1994,
Bouwmeester1997, Marcikic2003,Pan1998, Riedmatten2005}. There have
also been experiments of a BSA that detects all 4 Bell-states but
its overall efficiency does not reach 50\% and it requires the use
of an entangled ancilla photon-pair \cite{Walther2005}.

\section{Theory}
\subsection{Time-bin encoding}
In our experiments a qubit is encoded on photons using time-bins
\cite{Tittel1999}. This means that a photon is created that exists
in a superposition of two well defined instant in time (time-bins)
that have a fixed temporal separation of $\tau$. By convention the
Fock-state with $N=1$ corresponding to a photon in the early time
of existence $t_0$ is written as $\ket{0}$ and for the later time
$t_1=t_o+\tau$ as $\ket{1}$. Photons in such a state can be
created in several ways. The simplest method is to pass a single
photon through an unbalanced interferometer with a path length
difference of $nc\tau$, where $n$ is the refractive index. After
the interferometer the photon will be in the qubit state
$A\ket{0}+e^{i\alpha}B\ket{1}$. Here $A$ and $B$ are amplitudes
that depend on the characteristics of the interferometer and
$\alpha$ is the phase-difference between the interferometer paths
which is directly determined by $\alpha=\frac{2\pi
nc\tau}{\lambda} (\mbox{mod } 2\pi)$. For sake of readability we
will use the word \quote{qubit} when talking about a \quote{photon
that is in a qubit state}.

\subsection{Bell-state Analyzer}
In a large part of all experiments using Bell-state analyzers
(BSAs) that have been performed up to date, the BSA consists
essentially of a beamsplitter and single photon detectors (SPDs).
In such a beamsplitter-BSA(BS-BSA) the \quote{clicks} of the SPDs
are analyzed and, depending on their results, the input state will
be projected onto a particular Bell-state. With time-bin qubits as
described above a simple BS-BSA works as follows: two qubits
arrive at the same time on a beamsplitter but at
different entry ports. Since the 4 standard Bell-states %
\ba%
\ket{\phi_{\pm}}=\frac{1}{\sqrt{2}}(\ket{00}\pm\ket{11})\\
\ket{\psi_{\pm}}=\frac{1}{\sqrt{2}}(\ket{01}\pm\ket{10})
\ea%
form a complete basis we can write our 2-qubit input-state as a
superposition of these four states. One can calculate for each
Bell input-state the possible output states. These states can then
be detected using SPDs. The different detection patterns and their
probabilities are shown in Table \ref{table:simplecoincidances}.
\begin{table}
\begin{tabular}{l|c c c c c c c c c c c}
            D1& 00 &    & 11 &    & 01 &    & \hbox{ } 0 \hbox{ } & 0  & 1  & \hbox{ } 1 &\\
            D2&    & 00 &    & 11 &    & 01 & \hbox{ } 0 \hbox{ } & 1  & 0  & \hbox{ } 1 &\\ \hline
\ket{\phi_{+}}& 1/4& 1/4& 1/4& 1/4&    &    &    &    &    &    &\\
\ket{\phi_{-}}& 1/4& 1/4& 1/4& 1/4&    &    &    &    &    &    &\\
\ket{\psi_{+}}&    &    &    &    & \textbf{1/2}& \textbf{1/2}&    &    &    &    &\\
\ket{\psi_{-}}&    &    &    &    &    &    &    & \textbf{1/2}& \textbf{1/2}&    &\\
\end{tabular}
\caption{\label{table:simplecoincidances}The table shows the
probability to find specific coincidences as a function of the
input Bell-state in the case of a single beamsplitter as a BSA. A
\quote{0}(\quote{1}) in row D1 means that a photon was found at
detector \quote{D1} at time $t_0$($t_1$) etc. Note that only half
of the combinations of detection are possible for only one
Bell-state(the bold entries), therefore when such a combination is
found a projection onto this Bell-state was performed. The
theoretical success-probability is 50\%.}
\end{table}
By convention, a detection click at time \quote{0}(\quote{1})
means that the photon was detected in time-bin $t_0$($t_1$). The
output combinations show that, if one detects two photons in the
same path but in a different time-bin, the input state could only
have been caused by the Bell-state $\ket{\psi_+}$ and therefore
the overall state of the system is projected onto this state. When
the photons arrive at different detectors with a time-bin
difference the input-state is projected onto the state
$\ket{\psi_-}$. However, when one measures two photons in the same
time-bin in the same detector the state could either be
$\ket{\phi_+}$ or $\ket{\phi_-}$ and therefore the state has not
been projected onto a single Bell-state but onto a superposition
of two Bell-states. This method has a success rate of 50\% which
corresponds to the maximal possible success rate that can be
obtained while using only linear optics and no auxiliary photons
 \cite{Calsamiglia2001}.

\begin{figure}
\includegraphics[scale=0.3]{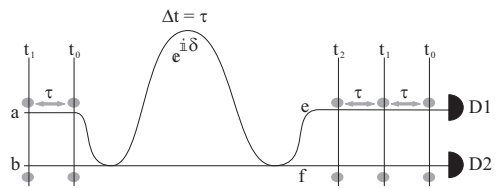}
\caption{\label{brunner}A schematic representation of the new type
of Bell-state Measurement. When two qubit states are sent into a
time-bin interferometer the output state is a mixture of photons
in two directional modes and three temporal modes. By looking at
certain combinations of these photons a Bell-state measurement can
be performed for three different Bell-states}
\end{figure}

Here we propose a new BSA which is capable of distinguishing more
than 2 Bell-states while still having the maximum success rate of
50\%. This is possible by replacing the beamsplitter with a
time-bin interferometer equivalent to the ones used to encode and
decode time-bin qubits \fig{brunner}. This BSA will be capable of
distinguishing three out of four Bell-states, but $\ket{\phi_+}$
and $\ket{\psi_-}$ will only de discriminated 50\% of the time as
will be explained shortly.
Two qubits enter in port $a$ and $b$, respectively. The first
beamsplitter acts like above, allowing the distinction of two
Bell-states ($\ket{\psi_+}$ and $\ket{\psi_-}$). A second
possibility for interference is added by another BS for which the
inputs are the outputs of the first BS, with one path having a
delay corresponding to the time-bin separation $\tau$. The
two-photon effects on this beamsplitter leads to fully
distinguishable photon combinations of one of the two remaining
Bell states ($\ket{\phi_+}$) while still allowing a partial
distinction of the first two.

One might expect that when it is possible to measure three out of
four states that the fourth, non-measured, state can simply be
inferred from a negative measurement result of the three
measurable states. This is, however, not the case. The above
described measurement is a POVM with $21$ possible outcomes, some
of these outcomes are only possible for one of the four input
Bell-states. Therefore when such an outcome is detected it
unambiguously discriminates the corresponding input Bell-state.
The rest of the $21$ outcomes correspond to outcomes which can
result from more than one input Bell-state. In other words, their
results are ambiguous and the input state is not projected onto a
single Bell-state but onto a superposition of Bell-states.

The state after the interferometer can be calculated for any
input-state using:
\ba%
\co{a}(t) \Rightarrow \frac{1}{\sqrt{4}}(&-\co{e}(t)+e^{i\delta}\co{e}(t+\tau)& \nonumber\\
&+i\co{f}(t)+ie^{i\delta}\co{f}(t+\tau)&) \\
\co{b}(t) \Rightarrow \frac{1}{\sqrt{4}}(&\co{f}(t)-e^{i\delta}\co{f}(t+\tau)& \nonumber\\
&+i\co{e}(t)+ie^{i\delta}\co{e}(t+\tau)&)
\ea%
where $\co{i}(j)$ is the creation operator of a photon at time $j$
in mode $i$. When the input-states are qubits and the photons are
detected after the interferometers the detection-patterns are
readily calculated and are shown in Table
\ref{table:coincidances}.
\begin{table*}
\begin{center}
\begin{tabular}{l|c c c c c c c c c c c c c c c c c c c c c c}
$D1$ & 00 &    & 11 &    & 22 &    & 01 &    & 02 &    & 12 &    & 0 & 2 & 1 & 1 & 0 & 1 & 2 & 0 & 2 \\
$D2$ &    & 00 &    & 11 &    & 22 &    & 01 &    & 02 &    & 12 & 0 & 2 & 1 & 0 & 1 & 2 & 1 & 2 & 0 \\
\hline \ket{\phi_{+}^\prime}
     &1/16&1/16&    &    &1/16&1/16&    &    &    &    &    &    &1/8&1/8&\textbf{1/2}&   &   &   &   &   &   \\
\ket{\phi_{-}^\prime}
     &1/16&1/16& 1/4& 1/4&1/16&1/16&    &    &    &    &    &    &1/8&1/8&   &   &   &   &   &   &   \\
\ket{\psi_{+}^\prime}
     &    &    &    &    &    &    & \textbf{1/8}& \textbf{1/8}&    &    & \textbf{1/8}& \textbf{1/8}&   &   &   &\textbf{1/8}&\textbf{1/8}&\textbf{1/8}&\textbf{1/8}&   &   \\
\ket{\psi_{-}^\prime}
     &    &    & 1/4& 1/4&    &    &    &    & \textbf{1/8}& \textbf{1/8}&    &    &   &   &   &   &   &   &   &\textbf{1/8}&\textbf{1/8}\\
\end{tabular}
\caption{\label{table:coincidances}The table shows the probability
to find any of the 21 possible coincidences as a function of the
input Bell-state. A \quote{0} in row D1 means that a photon was
found at detector \quote{D1} and at a time corresponding to the
photon having taken the short path in the interferometer and it
was originally a photon in time-bin $t_0$, a \quote{1} corresponds
to $t_0+\textbf{1}\times\tau$ with $\tau$ corresponding to a the
difference between the time-bins etc. Note that several
combinations of detection are possible for only one Bell-state(the
bold entries), therefore when such a combination is found a
Bell-state Measurement was performed. The theoretical probability
of a successful measurement is 0.5 which is the optimal value
using only linear optics \cite{Calsamiglia2001}.}
\end{center}
\end{table*}
The output coincidences on detectors $D1\mbox{ (port $e$)}$ and
$D2\mbox{ (port $f$)}$ are shown as a function of a Bell-state as
input. By convention, a detection at time \quote{0} means that the
photon was in time $t_0$ after the BSA-interferometer. This is
only possible if it took the short path in the BSA and it was
originally a photon in time-bin $t_0$ (Fig. \ref{brunner}).
Similarly a detection at time \quote{1} means that either the
photon was originally in $t_1$ and took the short path of the BSA
interferometer or it was in $t_0$ and took the long path. A
detection at time \quote{2} means the photon was in $t_1$ and took
the long path. In Table \ref{table:coincidances} we see that some
of the patterns corresponds to a single Bell-state and therefore
the measurement is unambiguous. For the other cases the result
could have been caused by two Bell-states, i.e. the result is
ambiguous and hence inconclusive. More specifically, the
Bell-state $\ket{\psi_{+}^\prime}$ is detected with probability 1,
$\ket{\phi_{-}^\prime}$ is never detected and both
$\ket{\psi_{-}^\prime}$ and $\ket{\phi_{+}^\prime}$ are detected
with probability 1/2.

The above described approach is correct in the case were the
separation $\tau$ of the incoming qubits is equal to the time-bin
separation caused by the interferometer. If this in not the case
and the interferometer creates a time-bin seperation of
$\tau+\frac{n\lambda\delta}{2\pi c}$, where $\delta$ is a phase,
the situation is slightly more complicated
In such a case, our BSA still distinguishes 3 Bell-states, but
these are no longer the standard Bell-states but are the
following:
\begin{eqnarray}
\ket{\phi^\prime_{\pm}} &=& \ket{00} \pm e^{2i\delta} \ket{11} = (\sigma_{\delta}\otimes\sigma_{\delta})\ket{\phi_{\pm}} \label{eq:phipm}\\
\ket{\psi^\prime_{\pm}} &=& e^{i\delta} (\ket{01} \pm \ket{10}) =
e^{i\delta}\ket{\psi_{\pm}} \label{eq:psipm}
\end{eqnarray}
Here $\sigma_{\delta}=P_{\ket{0}}+e^{i\delta}P_{\ket{1}}$ is a
phase shift of $\delta$ to be applied to the time bin $\ket{1}$.
These new Bell-states are equivalent to the standard states except
that the $\ket{1}$ is replaced by $e^{i\delta}\ket{1}$ for each of
the input modes.

\p In a realistic experimental environment the success
probabilities of the BSA are affected by detector limitations.
This is because existing photon detectors are not fast enough to
distinguish photons which follow each other closely (in our case
two photons separated by $\tau=1.2$ ns) in a single measurement
cycle. This limitation rises from the dead time of the
photodetectors. When including this limitation we find that the
maximal probabilities of success in our experimental setup are
reduced to $1/2$, $1/4$ and $1/2$ for $\psi_+$, $\psi_-$ and
$\phi_+$, respectively. This leads to an overall probability of
success of $5/16$ which is greater by $25\%$ than the success rate
of $1/4$ for a BSA consisting only of one beamsplitter and two
detectors with the same limitation. This limitation could be
partially eliminated by using a beamsplitter and two detectors in
order to detect the state 50\% of the time, or it could be
completely eliminated by using an ultra-fast optical switch
(sending each time-bin to a different detector). Both of these
methods are associated with a decrease in signal-to-noise ratio.
This is caused by additional noise from the added detector and by
additional losses from the optical switch, respectively.

\subsection{Bell-state analyzer for polarization qubits}
\begin{figure}
\includegraphics[width=4cm]{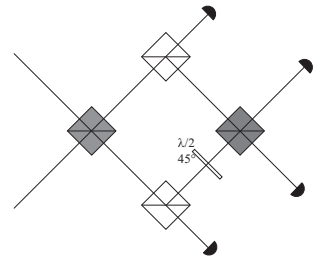}
\caption{\label{fig:polarisationBSA}A schematic representation of
the new type of Bell-state Analyzer for polarization qubits. The
grey cubes represent non-polarizing beamsplitters and the white
cubes are polarizing beamsplitters.}
\end{figure}

So far the discussion about this BSA only considered time-bin
qubits. The authors would like to note at this point that it is
also possible to implement a similar BSA for polarization encoded
photons. This can be done by the equivalent polarization setup as
shown in Fig. \ref{fig:polarisationBSA}. This setup would require
4 detectors but there will never be two photons on one detector
and therefore dead-times don't hinder the measurement of all the
detection patterns and the overall efficiency can reach 50\% with
today`s technology.

\subsection{4-Bell-state analyzer?} This paper discusses our
results testing a 3-Bell-state analyzer. It is obviously
interesting to also consider the possibility of a linear optics
4-Bell-state analyzer with 50\% efficiency and no ancilla photons.
Such a system was not used for the simple reason that there is no
known method to make such a measurement. Is there a fundamental
reason to suspect that such a BSA cannot be performed? No such
reason is known to the authors, therefor this article will be
limited to the 3-Bell-state analyzer.

\subsection{Teleportation}
One of the most stunning applications of a BSA is its use in the
teleportation protocol. In order to perform a teleportation
experiment an entangled qubit photon pair is created (EPR
\cite{Einstein1935a}) as well as a qubit to be teleported(Alice).
One photon of the entangled pair is made to interact with Alices
qubit in a BSA (Charly). This interaction followed by detection
projects the overall state onto a Bell-state (if the BSA is
successful). The remaining photon (Bob) now carries the same
information as the photon from Alice up to a unitary
transformation. The situation for the new BSA is slightly
different since the entangled pair is not a member of the detected
Bell-basis (Eq. \ref{eq:phipm} and \ref{eq:psipm}). However this
has no major influence on the theory. After a succesful
measurement of the BSA the remaining photon at Bob is equal to the
original qubit up to a unitary transformation. This transformation
however has to be adapted with regards to the standard case from
[$\openone$,$\sigma_z$,$\sigma_x$,$\sigma_x\sigma_z$] to
[$\sigma_{2\delta}^{-1}$,$\sigma_z\sigma_{2\delta}^{-1}$,$\sigma_x$,$\sigma_x\sigma_z$],
as can be seen from the following calculation.
\ba%
\ket{\zeta_{abc}}=\ket{\zeta}_a &\otimes &\ket{\phi_+}\\
= \frac{1}{2} (\ket{\phi^\prime_+}_{ab} &\otimes& \sigma_{2\delta}^{-1} \ket{\zeta}_c\\
+ \ket{\phi^\prime_-}_{ab} &\otimes& \sigma_z\sigma_{2\delta}^{-1} \ket{\zeta}_c \nonumber\\%
+ \ket{\psi^\prime_+}_{ab} &\otimes&  e^{-i\delta} \sigma_{x} \ket{\zeta}_c \nonumber\\%
+ \ket{\psi^\prime_-}_{ab} &\otimes& e^{-i\delta}\sigma_x\sigma_z \ket{\zeta}_c \nonumber%
\ea%
Recall
$\sigma_{2\delta}^{-1}=P_{\ket{0}}+e^{-2i\delta}P_{\ket{1}}$ is a
phase shift of the bit $\ket{1}$.

\section{Experimental Teleportation}
The new BSA was tested in a quantum teleportation experiment.
Presented in this section are the experimental setup that was used
as well as some of the required preliminary alignment experiments.
Finally the results of the experiment are given and discussed.

\subsection{Experimental setup}
\begin{figure*}
\includegraphics[scale=0.8]{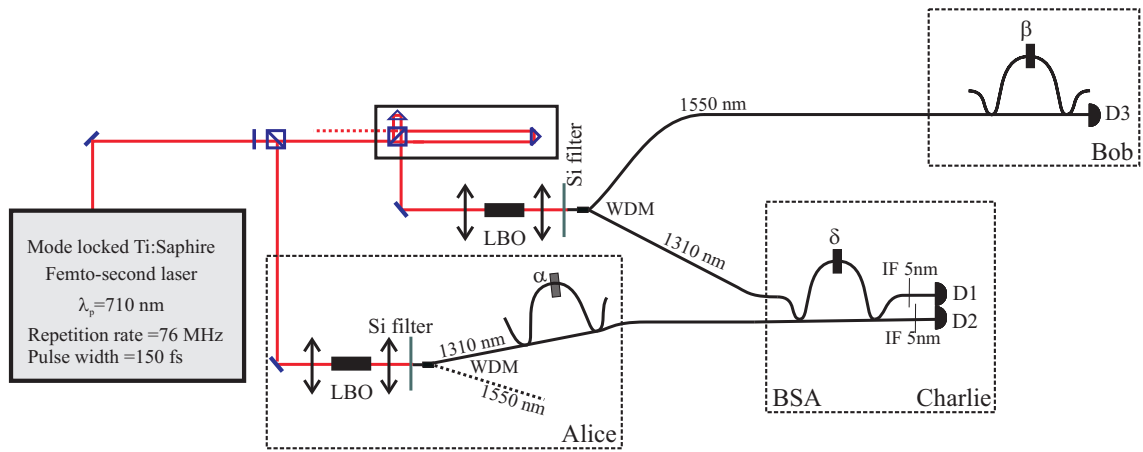}
\caption{\label{teleportbrunnerSETUP}A rough overview of the
experimental setup. The fiber interferometers shown
 here are in
reality Michelson-interferometers, for the interferometer in the
BSA two circulators are used to have two separate inputs and
outputs. Not shown in the figure is the method used for
stabilizing the interferometers.}
\end{figure*}

A rough schematic of the experimental setup is shown in Fig.
\ref{teleportbrunnerSETUP}, the experiment is an adaptation of a
setup used previously for long distance teleportation
\cite{Marcikic2003} and for entanglement swapping
\cite{Riedmatten2005}.
Alice prepares a photon in the state
$\ket{\zeta}_a=\frac{1}{\sqrt{2}}(\ket{0}+e^{i\alpha}\ket{1})$. A
BSA is used by Charly on Alices qubit combined with a part of an
entangled qubit pair. Bob analyzes the other half of the pair (the
teleported qubit) and measures interference fringes for each
successful BSA announced by Charly.

The setup consist of a mode-locked Ti-sapphire laser (MIRA with 8W
VERDI pump-laser, Coherent) creating 150fs pulses with a spectral
width of 4nm, a central wavelength of 711nm, a mean power of 400mW
and a repetition-rate of 80MHz. This beam is split in two beams
using a variable coupler ($\lambda/2$ and a PBS). The reflected
light (Alice) is sent to a scannable delay and afterwards to a
Lithium tri-Borate crystal(LBO, Crystal Laser) where by parametric
down-conversion a pair of photons is created at 1.31 and 1.55
$\mu$m. Pump light is suppressed with a Si-filter, and the created
photons are collected by a single mode optical fiber and separated
with a wavelength-division-multiplexer (WDM). The 1.55 $\mu$m
photon is ignored, whereas the photon at 1.31 $\mu$m is send to a
fiber interferometer which encodes the qubit state $\ket{\zeta}_a$
onto the photon.
The transmitted beam (Bob) is passed through an unbalanced
Michelson-type bulk interferometer. The seperation between the two
time-bins after this interferometer is considered as the reference
for all the other seperations. The phase of the interferometer can
therefore be considered as a reference phase and can be defined as
0. After the interferometer the beam passes a different LBO
crystal. The non-degenerate photons-pairs created in this crystal
are entangled and their state corresponds to
$\ket{\phi_+}=\frac{1}{\sqrt{2}}(\ket{00}+\ket{11})$.

The photons at 1.31$\mu$m are send to Charly in order to perform
the Bell-state measurement. To assure temporal
indistinguishability, Charly filters the received photons down to
a spectral width of 5nm using a bulk interference filter. Because
of this the coherence time of the generated photons is greater
than that of the photons in the pump beam, and as such no
distinguishablity between photons can be caused by jitter in their
creation time \cite{Riedmatten2003}. Bob filters his 1.55$\mu$m
photon to 15nm in order to avoid multi photon-pair events
\cite{Riedmatten2004, Scarani2005a}, this filtering is done by the
WDM that separates the photons at 1.31 and 1.5$\mu$m. This filter
is larger than Charlies filter for experimental reasons. A liquid
Nitrogen cooled Ge Avalanche-Photon-Detector (APD) D1 with passive
quenching detects one of the two photons in the BSA and triggers
the InGaAs APDs (id Quantique) D2 and D3. Events are analyzed with
a time to digital converter (TDC, Acam) and coincidences are
recorded on a computer.

Each interferometer is stabilized in temperature and for greater
stability an active feedback system adjusts the phase every $100$
seconds using reference lasers. The reference for Bob`s
interferometer is a laser (Dicos) stabilized on an atomic
transition at 1531nm and for both Alice`s and Charlies
interferometer a stabilized DFB-laser (Dicos) at 1552nm is used.
It is possible to use different lasers for Alice and Charly if one
wants to create two independent units. By using independent
interferometer units using different stabilization lasers it was
assured that this experiment is ready for use \quote{in the
field}. A more detailed description of the active stabilization of
the interferometers is given in ref \cite{Riedmatten2005}. For
sake of clarity the interferometers shown in the setup
\fig{teleportbrunnerSETUP} are Mach-Zender type interferometers
but in reality they are Michelson interferometers which use
Faraday-mirrors in order to avoid distinguishability due to
polarization differences \cite{Tittel2001}.

\subsection{Alignment experiments}
\begin{figure}
\begin{tabular}{l r}
\includegraphics[width=4cm]{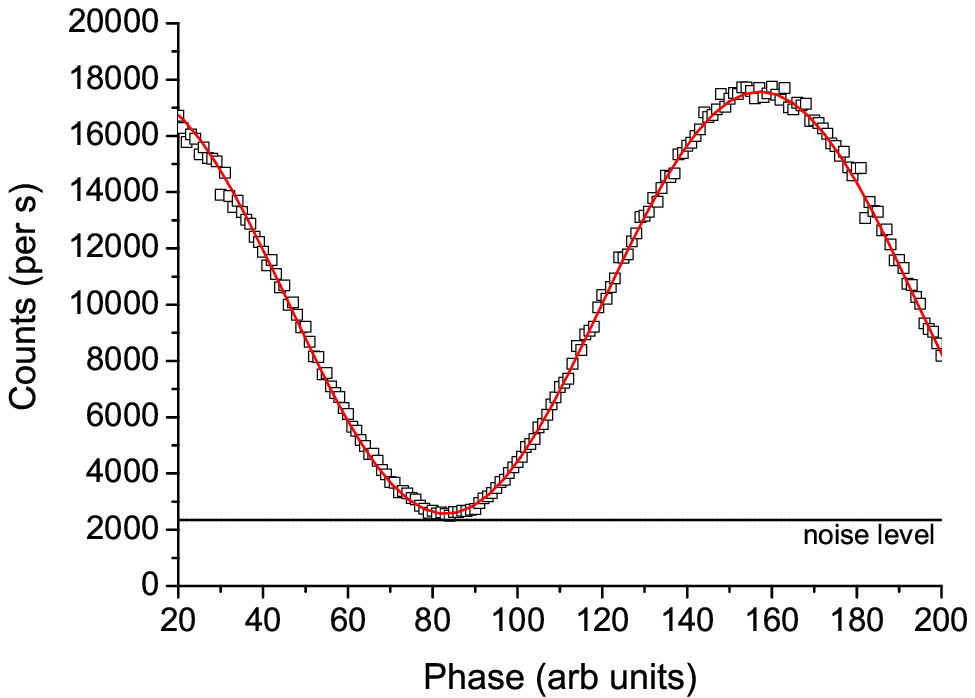}
&
\includegraphics[width=4cm]{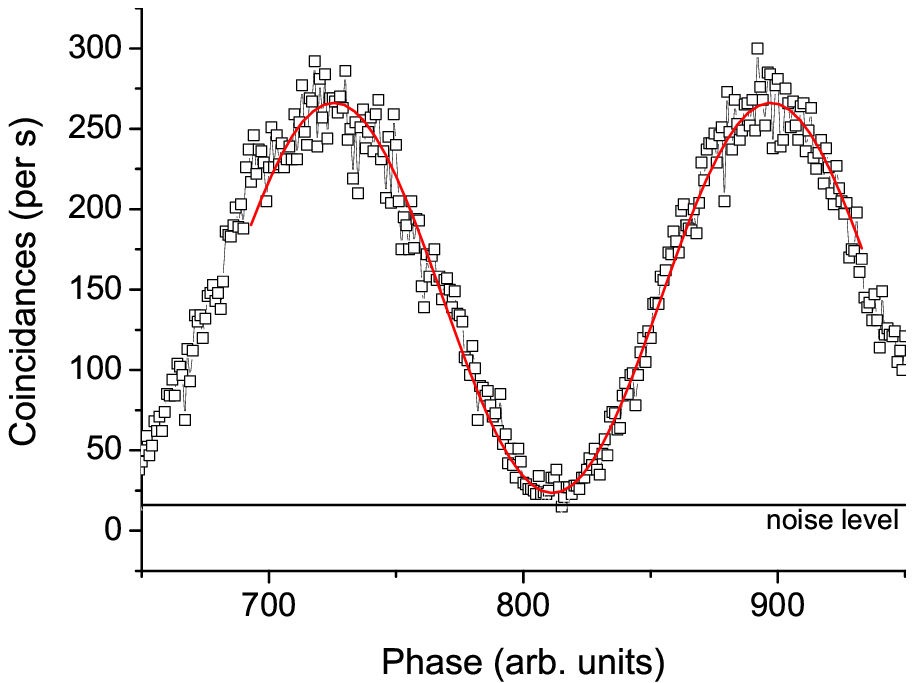}
\end{tabular}
\caption{\label{fig:alignmentexperiments}\emph{Left:} Single
photon interference with $V_{net}=0.97\pm0.01$. \emph{Right:}
Fransson-type Bell-test with $V_{net}=0.94\pm0.01$.}
\end{figure}

There are two important, non-trivial alignments that have to be
made before one can perform a quantum teleportation experiment
with time-bin encoded qubits. First, one would have to assure that
all the time-bin interferometers have the same difference in
length between the two paths. Second, it is required that there is
temporal indistinguishability between qubits coming from Alice and
Bob on the BSA. The equalization of the interferometers is needed
in order to assure that all the interferometers have a difference
in length of $\frac{c\tau}{n}$ with a precision higher than the
coherence length of the photons ($\approx$150$\mu m$). We have two
mechanisms to actively change the optical path lengths: the first
is changing the temperature of the interferometers and thus
allowing the long arm to increase/decrease its length more than
the short arm, the second is to directly change the length of only
one arm by means of a cylindric piezo-electric element. When
changing the voltage over the piezo we change the diameter of the
cylinder and thus the length of the fiber changes. This is used
for the active feedback stabilization system. In order to align
the interferometers with each other we perform two different
experiments: First, we optimize the visibilty of single photon
interference fringes for photons from Alice detected in $D1$. This
aligns Alices interferometer with the BSA interferometer. Next, we
optimize a Fransson-type Bell-test of the entangled photon-pair
\cite{Franson1989}. While optimizing this experiment we do not
change the BSA-interferometer. This optimization aligns the
bulk-interferometer and Bobs analysis interferometer to the other
two.
Using this method we found visibilities of $97\%\pm1\%$ for the
single photon interference and $94\%\pm1\%$ for the Bell-test
\fig{fig:alignmentexperiments}.

\begin{figure}
\begin{tabular}{l r}
\includegraphics[width=4cm]{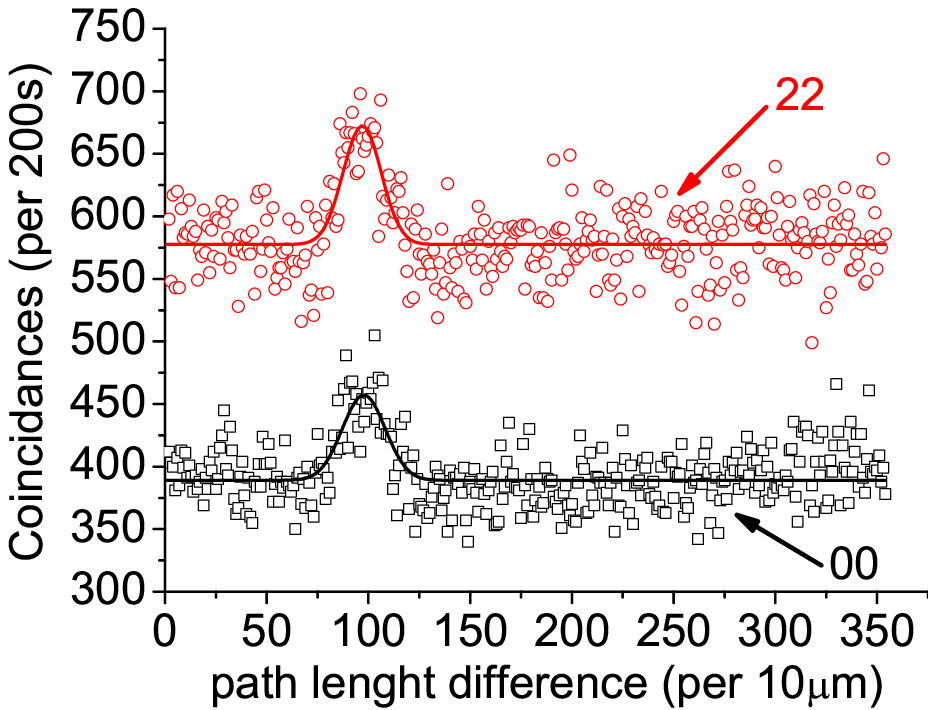} &
\includegraphics[width=4cm]{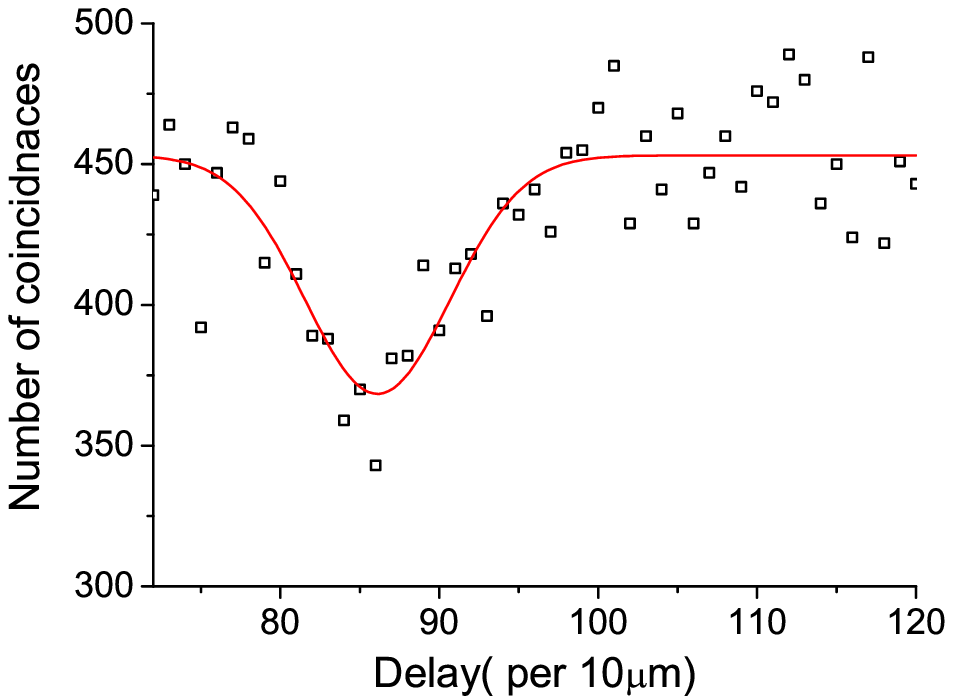}
\end{tabular}
\caption{\label{fig:antidip}\emph{Left, IF-BSA:} Graph showing the
number of measured coincidences as a function of a change in the
delayline. Both \quote{00} and \quote{22} clearly show an antidip
at the same location. The net visibilties are $V_{00}=32\pm3\%$
and $V_{22}=26\pm2\%$ \emph{Right, BS-BSA:} Graph showing the
decrease in the number of measured coincidences \quote{00} as a
function of a change in the delayline \cite{Hong1987}. The net
visibility is $V=29\pm3\%$.}
\end{figure}

The second alignment procedure is necessary in order to assure
temporal indistinguishability between the photons arriving at the
BSA. In the case of a BS-BSA this can be assured by performing a
Hong-Ou-Mandel dip type experiment \cite{Hong1987}, which is to
say, make a scan in a delay for one of the incoming photons and
look at a decrease in the number of coincidences as a result of
photon bunching \fig{fig:antidip}. The position where the minimum
is obtained corresponds to the point with maximal temporal overlap
of the two photons.

\begin{figure}
\includegraphics[width=8cm]{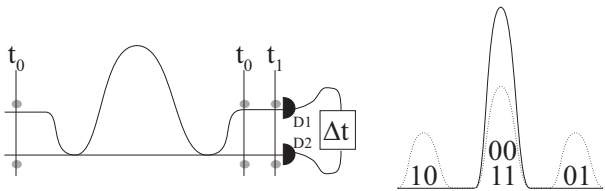}
\caption{\label{fig:alignmentantidip}The simple experiment on the
left
 (one photon in each input of an IF-BSA) will have the following property. If the photons are not
temporally indistinguishable one will find three different
coincidence peaks: \quote{10},\quote{00\&11} and \quote{01}
(dotted curve), however, if the photons are indistinguishable
there will be only one peak: \quote{00\&11}(plain curve). This is
caused by photon bunching.}
\end{figure}

In the case of an interferometer-BSA (IF-BSA) this procedure
becomes more complicated. We can no longer look at a mandel dip
because the second beamsplitter will probabilistically split up
the photons that bunched on the first beamsplitter. However the
photon bunching remains and it can still be seen by a different
method. Consider the situation where two single photons, both in
the state $\ket{0}$, are send to the different inputs of an IF-BSA
\fig{fig:alignmentantidip}. If the photons are not temporally
indistinguishable there are three possible output differences
between detection times, corresponding to \quote{10},
\quote{00\&11} and \quote{01}. If the photons are
indistinguishable they bunch at the first BS and therefore the
difference in arrival time between the photons has to be zero.
This means that \quote{10} and \quote{01} are not possible anymore
and the possibility for \quote{00\&11} is larger.
If the inputs are arbitrary qubits instead of the simple example
above there will be more coincidence possibilities and some of
them will be subject to single-photon interference and/or photon
bunching. It is possible to see an increase in the coincidences
for \quote{00} and \quote{22}, which is not affected by
single-photon interference, for similar reasons as the increase
that was explained above. These coincidences can be measured in a
straightforward way with our setup. A more rigorous calculation
and explication of this alignment procedure is given in the
appendix. A typical result of an experiment in which the count
rate is measured while changing a delay is shown in Fig.
\ref{fig:antidip} and clearly shows the expected increase in
 count rate.

The measured antidips have a net visibility of $32\pm3\%$ and
$26\pm2\%$ after noise substraction. The maximal attainable value
is $1/3$ due to undesired but unavoidable double-pair events (see
appendix). The large visibilities mean that the temporal
indistinguishability is very good, this will thus not be limiting
for our experiments. The noise substraction for this estimation is
justified because in a teleportation experiment the noise will be
reduced since one will consider only 3-photon events. The
difference in height of the two coincidences is related to an
electronic loss of signal in an electrical delay line.

\subsection{Experimental Results}
\begin{figure}
\includegraphics[width=7cm]{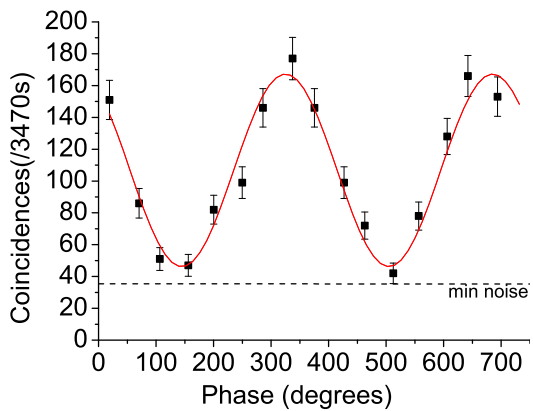}
\caption{\label{fig:telepclone}The Result of the 1-Bell state
teleportation experiment (a beamsplitter instead of the
interferometer) with $F_{raw}=0.79\pm0.02$ and
$F_{net}=0.91\pm0.02$. }
\end{figure}

Two different types of teleportation experiments were performed.
Namely a standard BS-BSA teleportation in order to benchmark our
equipment followed by the new IF-BSA experiment. For the BS-BSA
the main difference with regards to previous experiments \cite{
Riedmatten2004, Riedmatten2005} was that the interferometers now
all had an active stabilization. This allows for large stability
and long measurement times. The experiment consisted of Bob
scanning of his interferometer phase while the other
interferometers where kept constant, we therefore expect to find
an interference curve of the form $1+V sin(\beta+\alpha)$ where
$\alpha$ is kept constant. The results of the experiment
\fig{fig:telepclone} clearly shows the expected behavior.
The visibility measured was $V=0.57\pm0.03$ ($F=0.79\pm0.02$).
After conservative noise substraction we find $V=0.83\pm0.04$
($F=0.91\pm0.02$). This clearly is higher than the strictest
threshold that has been associated with quantum teleportation of
$F=5/6$ \cite{Grosshans2001,Bruss1998a}. The limiting factors of
this experiment are the detectors and the fiber-coupling after the
LBO crystals.

After this experiment  the setup was changed to the IF-BSA. The
count-rates in this experiment with regards to the previous one is
reduced due to two reasons. The introduction of the
BSA-interferometer and its stabilization optics means an
additional 3dB of 
loss which reduces count rates. Another difference is that now the
counts are distributed over three different Bell-states whereas
before there was only one. Therefore an overall reduction of
counts per state will occur. Combined these effects result in a
large reduction of the count rate per Bell-state. This problem was
overcome by, on one hand, an overall increase of the BSA
efficiency by 1/4 (from 25\% for the BS-BSA to 31.25\% for the
IF-BSA) and, on the other hand, by integrating data over longer
time periods. During the teleportation-experiment scans were made
in the interferometer of Alice rather than Bob. This was done
since the most important noise is dependant of the phase of Bob`s
interferometer but not of Alice`s (more details are given in the
next subsection). The experiments were performed with
approximately 4.4 hours per phase setting in order to have low
statistical noise.

For this IF-BSA all the different unambiguous results (Table
\ref{table:coincidances}) where analyzed both separately (for
example \quote{02}) and combined as a Bell-state (for example
$\ket{\psi_-}=$\quote{02}$+$\quote{20}). For the separate results
it is expected that each BSA outcome will have count rates
depending on the phases of the interferometers as $R(1+V
cos(\alpha+\rho))$. Here $R$ is dependant of the overall
efficiency of the experiment and is different for each BSA outcome
and $\rho$ is a combination of the constant phases of the
interferometers of Bob and Charlie and is different for different
BSA results:
\ba%
\ket{\psi_+},\ket{01},\ket{10},\ket{12},\ket{21}  &\rightarrow& \rho=\beta \\
\ket{\phi_+},\ket{11} &\rightarrow& \rho=-\beta-2\delta \\
\ket{\psi_-},\ket{02},\ket{20} &\rightarrow& \rho=\beta+\pi
\ea%
As is evident from the differences in $\rho$ we expect that
fringes corresponding to one particular Bell-state are in phase
with each other, but have a well determined phase-difference with
fringes corresponding to another Bell-state.

\begin{figure}
\begin{tabular}{l r}
\includegraphics[width=4cm]{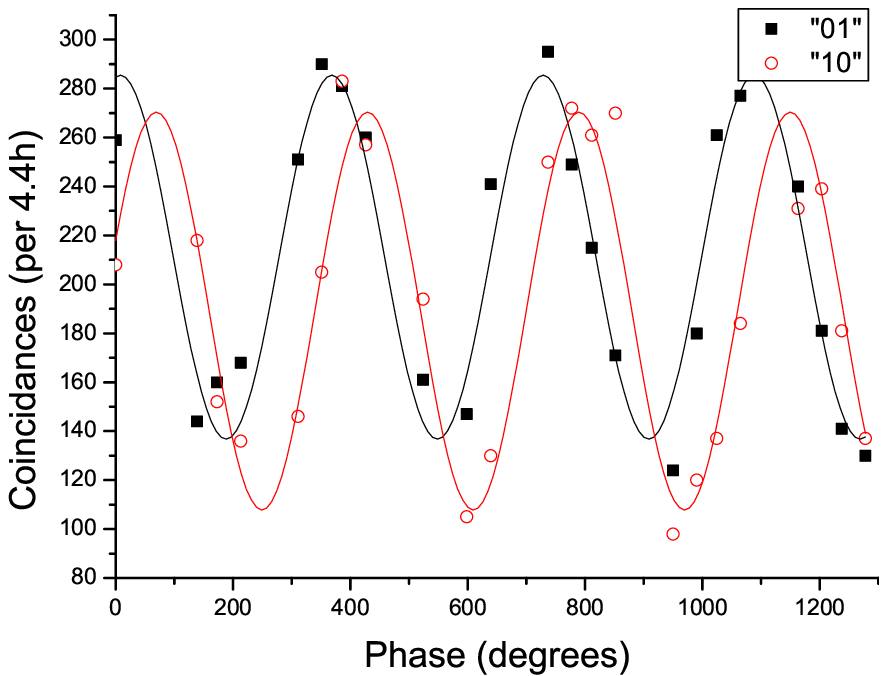}
& \includegraphics[width=4cm]{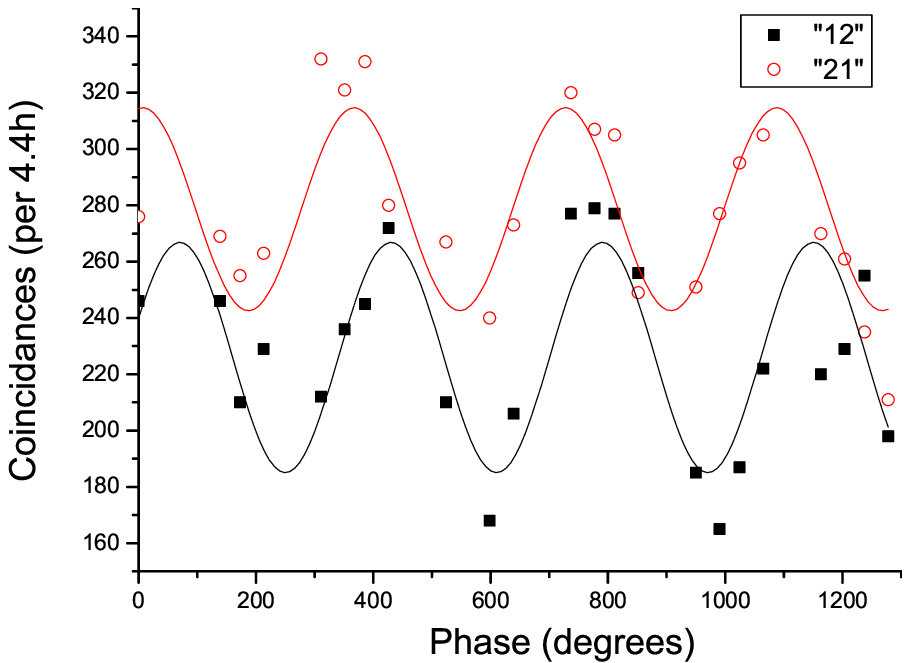}\\
\includegraphics[width=4cm]{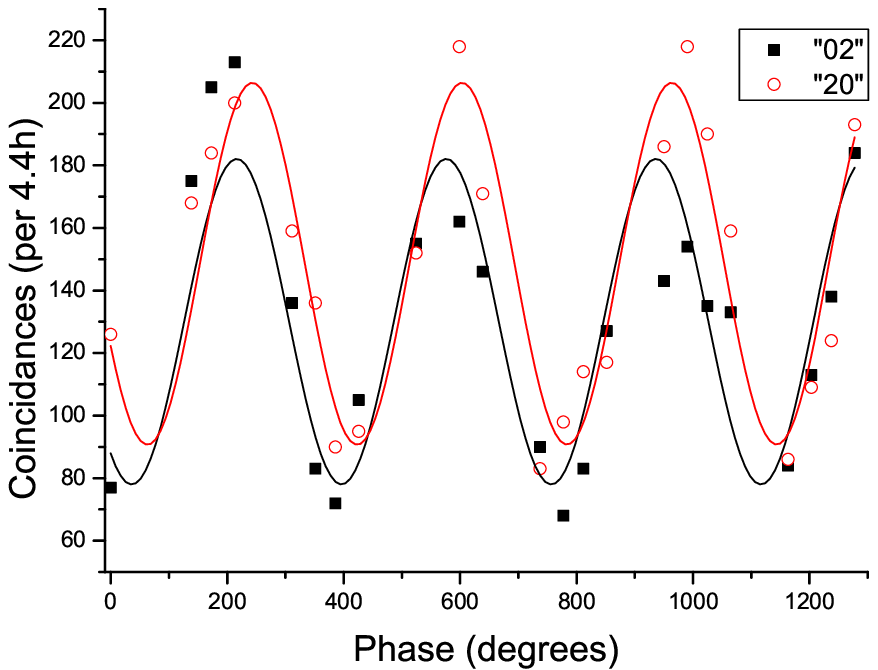}
&\includegraphics[width=4cm]{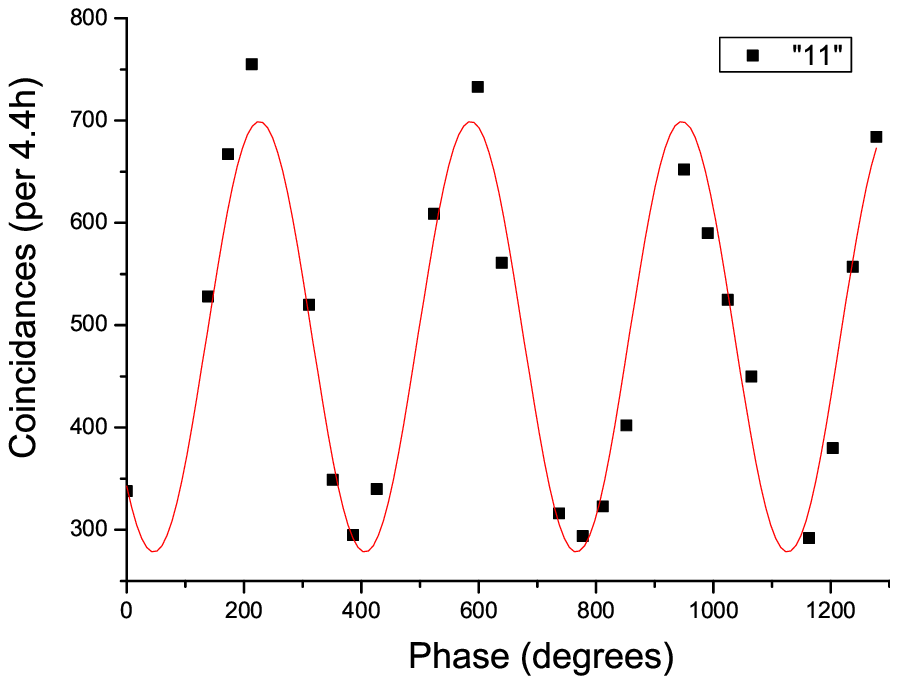}
\end{tabular}
\caption{\label{fig:individualvis}Measured coindance counts as a
function of phase. \emph{Top-Left:} BSA-results \quote{01},
\quote{10}. \emph{Top-Right:} BSA-results \quote{12}, \quote{21}.
\emph{Bottom-Left:} BSA-results \quote{02} and \quote{20}.
\emph{Bottom-Right:} BSA-results \quote{11}. Note that for each
graph there is an unknown, but equal, offset of the phase-value.}
\end{figure}

The measured count rates as a function of the phase of Alices
interferometer are shown in \fig{fig:individualvis}.
Note that, due to experimental restrictions, the absolute phases
of the interferometers are not known and therefor all phase-values
have an unknown offset. The results clearly show that each of the
outcomes has the expected interference behavior. Furthermore, the
fringe corresponding to \quote{01} is in phase with the fringe
\quote{21}. The same is true for the fringe \quote{10} with
\quote{12} and for \quote{02} with \quote{20}. It is expected that
all four of the fringes corresponding to \ket{\psi_+}
(\quote{01},\quote{10},\quote{12} and \quote{21}) are all in phase
with each other, but there is a clear phase-shift between the
first two and the last two. The average phase of these four
fringes is different by $180^o$ from the fringes corresponding to
\quote{02} and \quote{20} as expected. The fringe corresponding to
\quote{11} is in phase with the fringes of \quote{02} and
\quote{20} as was expected since for this measurement we had
arranged $2(\beta+\delta)=\pi(\mbox{mod} 2\pi)$. The results of
the fits to these fringes is shown in Table
\ref{table:individualvis}. The differences in phase and visibility
are in part due to noise (see next subsection)
\begin{table}
\begin{center}
\begin{tabular}{l r|c c|c c|c c}
\multicolumn{2}{c|}{Result 3BSA} & $V_{raw}$ & $V_{net}$ & $\rho_{raw}$ & $\rho_{net}$ & $P_{raw}$ & $P_{net}$\\\hline%
\multirow{4}{*}{\ket{\psi_{+}}}&\ket{01}&35$\pm$ 3 &61$\pm$ 6&278 $\pm$ 4&279$\pm$5&13$\pm$1&14$\pm$1\\%
&\ket{10}&43$\pm$ 3 &72$\pm$ 13&339 $\pm$ 4&338$\pm$8&11$\pm$1&13$\pm$1\\
&\ket{12}&18$\pm$ 3 &64$\pm$ 7&340 $\pm$ 7&340$\pm$4&14$\pm$1&7$\pm$1\\
&\ket{21}&13$\pm$ 2 &36$\pm$ 2& 227 $\pm$ 9&278$\pm$1&17$\pm$1&10$\pm$1\\\hline%
\ket{\phi_{+}}&\ket{11}&43$\pm$ 3 &55$\pm$ 2&136 $\pm$ 3&136$\pm$10&29$\pm$1&41$\pm$1\\\hline%
\multirow{2}{*}{\ket{\psi_{-}}}&\ket{02}&40$\pm$ 5 &83$\pm$ 13&126 $\pm$ 12&126$\pm$8&8$\pm$1&6$\pm$1\\
&\ket{20}&39$\pm$ 4 &62$\pm$ 10&153 $\pm$ 4&153$\pm$8&9$\pm$1&9$\pm$1\\
\hline\hline
\multicolumn{2}{c|}{\ket{\psi_{+}}}&$22\pm 1$& $51\pm 3$  & 311 $\pm$ 3& 311$\pm$ 3&54$\pm$1&43$\pm$1\\
\multicolumn{2}{c|}{\ket{\phi_{+}}}&$43\pm 3$& $55\pm 2$ & 136 $\pm$ 10& 136$\pm$ 3&29$\pm$1&41$\pm$1\\
\multicolumn{2}{c|}{\ket{\psi_{-}}}&$38\pm 5$& $69\pm 10$  & 140 $\pm$ 6& 140$\pm$ 7&17$\pm$1&15$\pm$1\\
\end{tabular}
\caption{\label{table:individualvis} For each of the different
detection possibilities the fitted result are shown before and
after noise correction.\quote{$V$} refers  to the Visibility (\%),
\quote{$\rho$} to the phase of the fringes (degrees, unknown
offset) and \quote{P} to the normalized probabilities of a
coincidence detection (\%) . The last three rows correspond to
fits made after adding the corresponding BSA outcomes, therefore
these values can be slightly different from the average of the
individual results.}
\end{center}
\end{table}

The results corresponding to each of the three possible
Bell-states can be found by adding the measurements of the
constituent parts. When doing this one would expect coincidence
fringes of the form $R(1+V cos(\alpha+\rho))$ with $R$ and $\rho$
as above. This corresponds to three distinct interference fringes,
with \ket{\phi_+} and \ket{\psi_-} in phase and \ket{\psi_+} with
a $180^o$ phase-difference with respect to the other two.

\begin{figure}
\includegraphics[width=7.5cm]{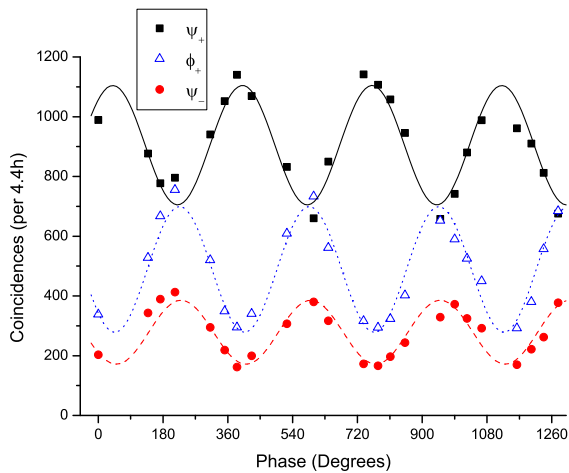}
\caption{\label{results} Uncorrected teleportation fringes found
when scanning the interferometer at Bob. The fitted curves have
visibilities of 0.22,0.43 and 0.38 for $\ket{\psi_+}$,
$\ket{\phi_+}$ and $\ket{\psi_-}$. The average visibility of the
BSA is $V_{avg}=0.34$ (F=0.67).}
\end{figure}

\begin{figure}[t]
\includegraphics[width=7.5cm]{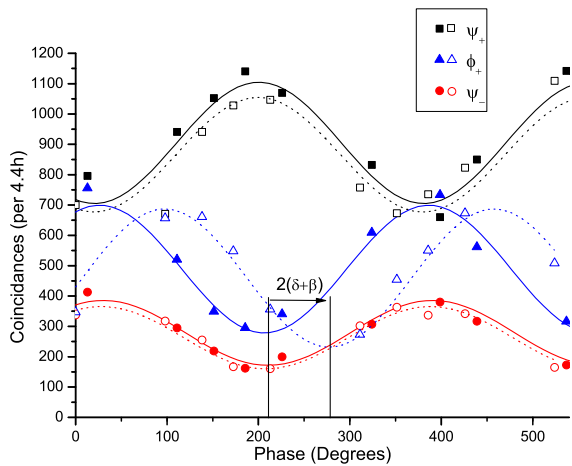}
\caption{\label{resultsdelta}Teleportation fringes measured in two
distinct measurement with a $\delta$ which had changed by $70 \pm
10^o$. In the measurement a clear shift is visible of the fringe
$\ket{\phi_+}$ by $74^o$ with regards to the other fringes.}
\end{figure}

In Fig. \ref{results} we show the raw coincidence interference
fringes between detection rate at Bob and a successful BSA as a
function of a change of phases in Alices interferometer. As
expected fringes for $\ket{\psi_-}$ and $\ket{\psi_+}$ have a
$180^o$ phase difference due to the phase flip caused by the
teleportation. On the other hand the fringe for $\ket{\phi_+}$ is
in phase with the $\ket{\psi_+}$ as expected. The raw visibilities
obtained for the projection on each Bell-state are
$V_{\psi_-}=0.38\pm0.05$, $V_{\psi_+}=0.22\pm0.01$,
$V_{\phi_+}=0.43\pm0.03$ which leads to an overall value of
$V=0.34\pm0.06$ ($F=0.67\pm3$). In order to check the dependence
of $\ket{\phi_+}$ on $\delta$ we also performed a teleportation
with a different value for $\delta$ and we clearly observe the
expected shift in the fringe \fig{resultsdelta} while measuring
similar visibilities.

Note that Bob is able to derive the phase value $\delta$ of the
BSA interferometer just by looking at the phase differences
between the fringes made by $\psi_{\pm}$ and $\phi_{+}$ and his
knowledge about $\beta$. It is important to know $\delta$ since
this allows Bob to perform the unitary transformation
$\sigma_{2\delta}$ on the teleported photon.

Since the count rates were quite low we expected to have an
important noisefactor, an analysis of the noise follows in the
next subsection.

\subsection{Noise Analysis and Discussion}
In the case of the BS-BSA the noise analysis is straightforward.
All of the important noise counts are completely independent of
the phase, since  they concern situations in which there is no
single-photon interference possible. The most important sources of
noise were estimated and then measured. The estimated SNR was
$2.6$, measurements find a SNR of approximately $2.2\pm0.5$. The
largest source of noise are darkcounts at one detector combined
with two real detections.

\begin{figure}[]
\begin{tabular}{l r}
\includegraphics[width=4cm]{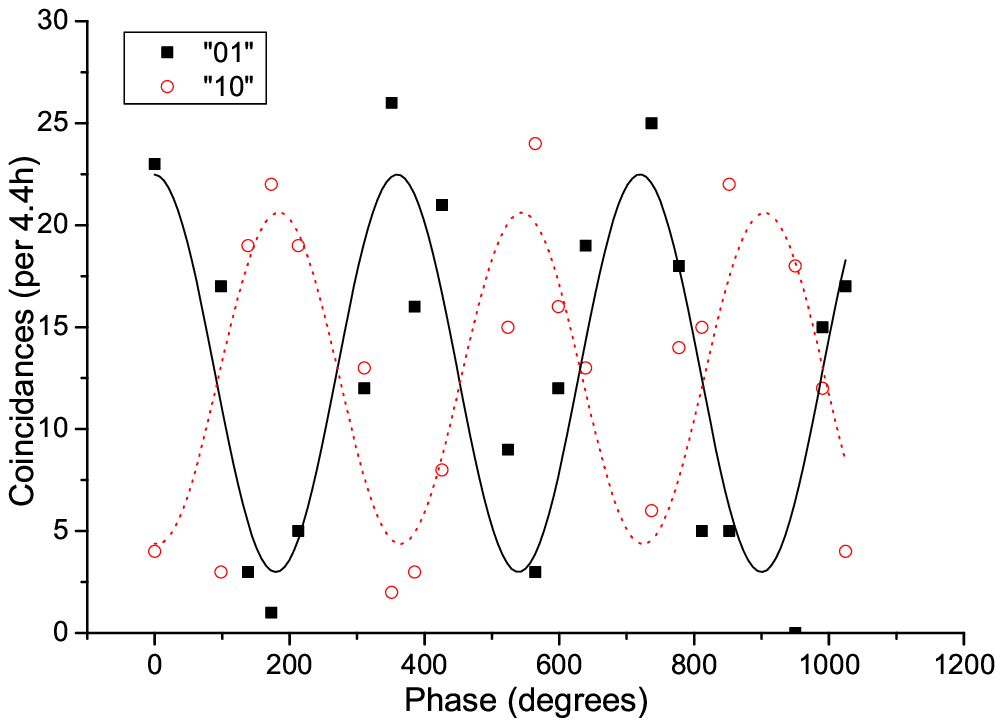} &
\includegraphics[width=4cm]{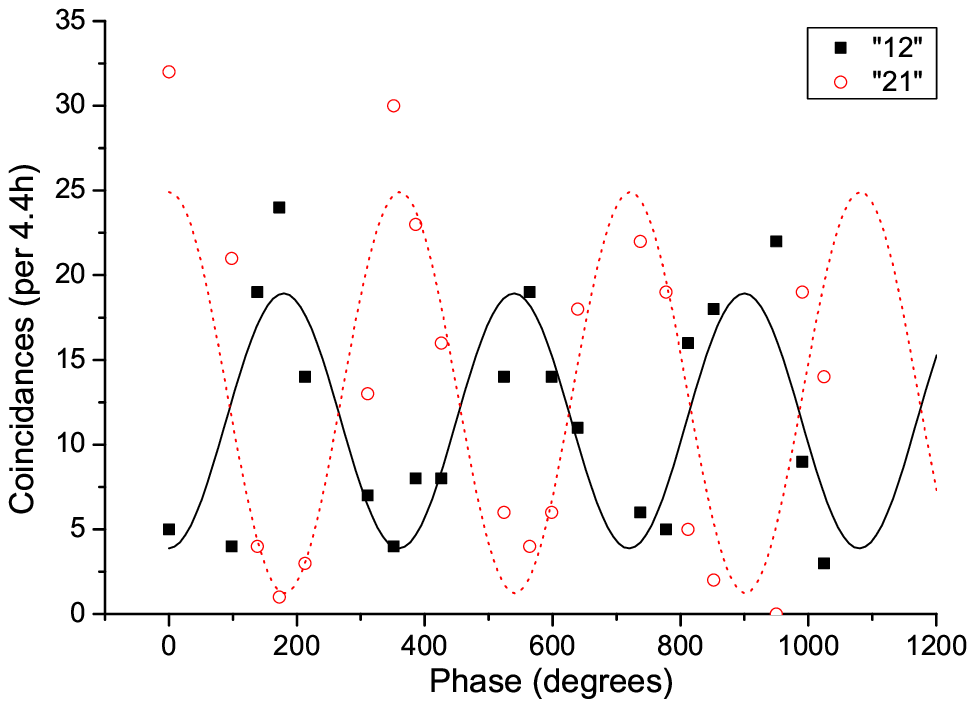}
\end{tabular}
\caption{\label{fig:noisefluctuations}Measurements of the Noise
for a interferometer-BSA teleportation experiment. \emph{Left:}
\quote{01} and \quote{10} are in antiphase as expected and have
Visibilities of $V_{01}=0.77\pm0.12$, $V_{10}=0.65\pm0.12$
\emph{Right:} \quote{12} and \quote{21} have a $\pi$ phaseshift as
expected and have Visibilities of $V_{12}=0.66\pm0.14$,
$V_{21}=0.91\pm0.13$}
\end{figure}

The situation for the IF-BSA is more complicated. The additional
interferometer has an unfortunate side-effect. There are now
possibilities for noise to depend on the phases of the
interferometers. In other words, while measuring interference
fringes there are also noise fringes. It it obviously important to
be able to distinguish between the two. The most important
fluctuating noise is caused by false coincidence-detections that
involve one (or more) photons coming from Alice and no photons
from the EPR-source at the BSA. These noise sources depend on the
phases $\alpha$ and $\delta$ since the photon coming from Alice
experiences single-photon interference. Since during the
experiment $\alpha$ is changed the noise-rate also changes. The
period of this change is the same as for teleportation, however,
there is a $\pi$ phaseshift between \quote{01}(\quote{21}) and
\quote{10}(\quote{12}) that is not present in the teleportation
signal. Such a noise influences the results of our measurements in
different ways, first of all, the visibilities are altered and are
smaller for \quote{01} and \quote{21} but larger for \quote{10}
and \quote{12}. Secondly, when the fringe is not in phase with the
teleportation signal there is a phaseshift in the opposite
direction for \quote{01} and \quote{21} with regards to \quote{10}
and \quote{12}. In our measurements the phases were arranged in
such a way that this second effect would not take place since the
fringes would be completely in or out of phase with the
teleportation signal. This noise was measured and the result
\fig{fig:noisefluctuations} clearly shows the expected fringes and
phaseshifts.

Other possibilities for fluctuating noise sources are when no
photons from Alice arrive at the BSA. In this case single (or
multiple) photon-pairs from the EPR-source combined with
darkcounts will give coincidences that depend on the phases
$\beta$ and $\delta$. This corresponds to a combination of a
Fransson-type Bell-test with a darkcount. The fluctuation of this
noise was avoided in our experiment since we only changed the
phase of Alices interferometer($\alpha$).

Not all possible sources of noise depend on the phases of the
interferometers, there are also stable sources of noise, which are
different for each of the BSA-possibilities. The average value of
the most important noise sources are shown in table
\ref{table:noise}, which shows that by choosing to scan Alice
instead of Bob a large fluctuating noise was avoided.
\begin{table*}
\begin{tabular}{l|ccc|ccc|ccc|ccc|ccc|ccc|ccc}
 &  & 01 &  &  & 02 &  &  & 10 &  &  & 11 &  &  & 12 &  &  & 20 &  &  & 21 &  \\ \hline
Source Alice Blocked & 70 & $\pm$ & 4 & 60 & $\pm$ & 4 & 60 & $\pm$ & 4 & 88 & $\pm$ & 6 & 147 & $\pm$ & 6 & 46 & $\pm$ & 3 & 157 & $\pm$ & 5 \\
EPR to Bob Blocked & 2.7 & $\pm$ & 0.1 & 1.0 & $\pm$ & 0.1 & 3.5 & $\pm$ & 0.1 & 5.3 & $\pm$ & 0.1 & 4.3 & $\pm$ & 0.1 & 1.7 & $\pm$ & 0.1 & 4.2 & $\pm$ & 0.1 \\
EPR to BSM Blocked & 13 & $\pm$ & 1 & 7 & $\pm$ & 1 & 12 & $\pm$ & 1 & 15 & $\pm$ & 1 & 11 & $\pm$ & 1 & 7.8 & $\pm$ & 1 & 13.07 & $\pm$ & 1 \\
\end{tabular}
\caption{\label{table:noise} The average noisecounts of several
noise possibilities. Note that each measured value concerns a
combination of different sources of noise. The most important
noise (source Alice blocked) didn't fluctuate during the
experiment because the scan in phase was done by Alice.}
\end{table*}
It also shows the fluctuating noise from Alice is only a small
part of the total noise and therefore its effect will only be
limited. Another source of errors that is different for each
coincidence combination is electronical loss. These losses are
caused by long (up to 100 ns) electronical delays that are
required in the treatment of the coincidence signals.

\begin{figure}[]
\includegraphics[width=8cm]{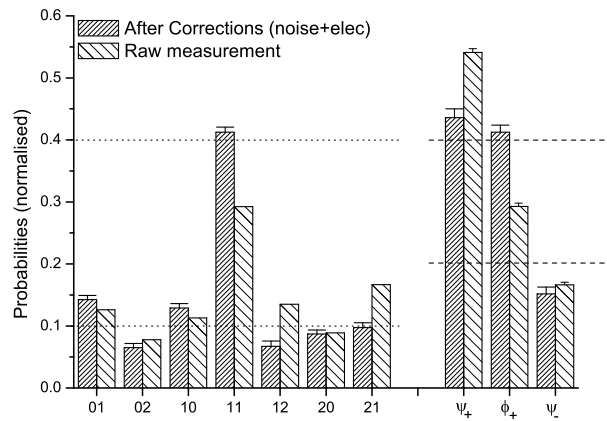}
\caption{\label{fig:proba} Normalized probabilities of detecting a
coincidence or a Bell-state. The expected value for the
coincidence \quote{11} is $0.4$ and for the other coincidances
$0.1$. For the Bell-state \ket{\psi_-} one expects $0.2$ and $0.4$
for \ket{\psi_+} and \ket{\phi_+}.}
\end{figure}

The results, after noise-substraction and correction for
electronical transmission differences for the individual
coincidence combinations, are shown in table
\ref{table:individualvis}. There is a clear agreement with theory,
for example the probability of finding a \quote{11} is
approximately $4$ times larger than the probability for any of the
other possibilities \fig{fig:proba}.

There are a few noteworthy differences between the results and
theory. First of all there are small differences in visibility,
these are probably caused by several small unmeasured noise
sources and partially they are real physical differences which are
caused by imperfect interferometers, an indication of these
imperfections is given by the quality of the alignment
experiments. Second the phases of the curves show an interesting
phase difference between \quote{01}(\quote{12}) and
\quote{10}(\quote{21}). The reason for this shift is unknown, but
the average value of the two phases corresponds with the phase
that is expected from the curve for \quote{02} and \quote{20}.
This suggests that this effect is caused by a fluctuating noise
that is out of phase with the teleportation fringe.

\begin{figure}[]
\includegraphics[width=8cm]{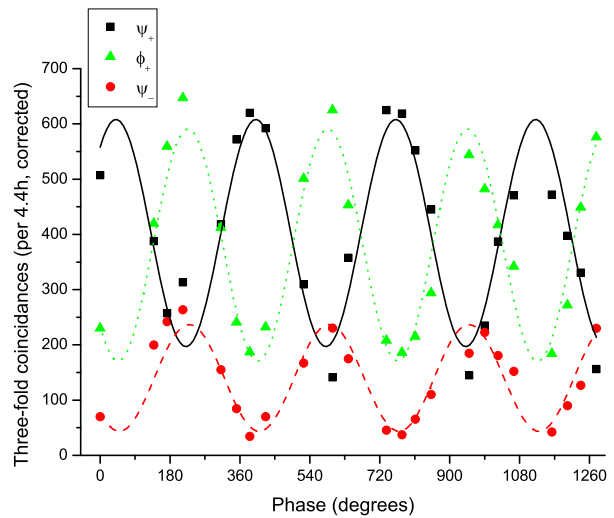}
\caption{\label{fig:correctedteleportation}Result of the
Teleportation experiment using an interferometer Bell-state
analyzer. These results are corrected for noise and electronic
differences.}
\end{figure}

When the different possibilities of the BSA are summed, in order
to have the Bell-states, the noise will no longer have any
fluctuations. This is because the different noise-possibilities
had a $\pi$ phase difference. After summing the different parts of
the noise of a Bell-state the result will be constant. For
example, the noise of \quote{01} combined with \quote{10} is
approximately constant. The overall resulting noise is in practise
independent of the phase. The results after noise substraction and
correction for electronic transmission differences are shown in
Fig \ref{fig:correctedteleportation} and Table
\ref{table:individualvis} .
The results show excellent correspondence between theory and
experiment. The visibilities are similar within their errors. The
difference in phase between \ket{\psi_+} and \ket{\psi_-}
($189^o\pm9^o$) corresponds with theory ($180^o$). Also, since the
phases were arranged in such a way that $\beta=-\delta (\mbox{mod
} 2\pi)$, the fringe of \ket{\phi_+} is in phase with \ket{\psi_-}
(phase difference of $4^o\pm9^o$). The normalized probabilities of
a measurement \fig{fig:proba} show that \ket{\psi_+} and
\ket{\phi_+} have the same probability (43\% resp. 41\%) and these
values correspond with the theoretical value of 40\%. The
probability of \ket{\psi_-} is 15\% with a theoretical value of
20\%. These excellent agreements with theory suggest that the
discrepancies as seen for the individual results are caused by
differences in noise that cancel out when they are added to each
other.

\section{Conclusions}
In conclusion we have shown experimentally that it is possible to
perform a three-state Bell analysis while using only linear optics
without the use of ancilla photons. In principle this measurement
can reach a success rate of 50\%. We have shown some of the
techniques that have to be used to align a teleportation
experiment which uses this BSA. Our teleportation experiment shows
a non-corrected overall fidelity of F=$67\%$, after noise
substraction we find F=$76\%$. Also, we performed a teleportation
experiment with a one state BSA which exceeded the cloning limit.
\p The authors acknowledge financial support from the European
IST-FET project "RamboQ", the european project "QAP" and from the
Swiss NCCR project "Quantum Photonics" and thank C. Barreiro and
J.-D. Gautier for technical support. Also we would like to express
our gratitude to Dr. Keller of the ETH Zurich for the lending of
materials.

\section{\label{app:antidip}Appendix: Temporal Alignment}

For a BSA to work it is important to have complete
indistinguishability of the incoming qubits. This includes a
indistinguishability in time. In order to align the path lengths
in an experiment it is useful to perform photon bunching
experiments, since photon bunching only occurs for
indistinguishable photons. In the case of a teleportation
experiment using a BS-BSA it is possible to perform a Mandel-dip
experiment \cite{Hong1987} by looking at the coincidence-rate
\quote{00} or \quote{11}. A decrease in the number of coincidences
between the BSA-detectors is observed when the photons from Alices
source and the EPR-source are indistinguishable. When an IF-BSA is
used it is not trivial to directly measure such an effect, without
having to make significant changes to the optical setup (such as
replacing the interferometer by a beamsplitter) in between two
teleportation experiments. In order to avoid any changes to the
setup another method of checking indistinguishability was used. An
increase in the number of coincidences \quote{00} or \quote{22} is
dependant on indistinguishability, as was explained in the main
text of the article. The difference is clearly seen by calculating
the probability to find \quote{0} in both D0 and D1 for
indistinguishable photons ($P(\mquote{00}|\mbox{aligned})$) and
distinguishable ($P(\mquote{00}|\mbox{non-aligned})$) photons:
\ba%
\nonumber\mbox{1 photon from both sources}\\
P(\mquote{00}|\mbox{aligned})&=&1/4\\
P(\mquote{00}|\mbox{non-aligned})&=&1/8
\ea%

The maximum visibility when measuring the difference between
aligned and non-aligned can be calculated by taking into account
the probability to create two photons in Alice
$(P(\mquote{00}|(a^\dagger)^2)$ or at the EPR-source
$P(\mquote{00}|(b^\dagger)^2)$.
\ba%
P(\mquote{00}|(a^\dagger)^2)&=&P(\mquote{00}|(b^\dagger)^2)=1/8\\
V&=&-\frac{P_{out}-P_{in}}{P_{out}}\\
P_{out}&=&1/8+2*1/8=3/8\\
P_{in}&=&2/8+2*1/8=4/8\\
\nonumber V_{max}&=&1/3
\ea%
Note that when making measurements of \quote{antidips} the photons
at Bob are completely ignored.

\begin{figure}[]
\begin{tabular}{l r}
\includegraphics[width=4cm]{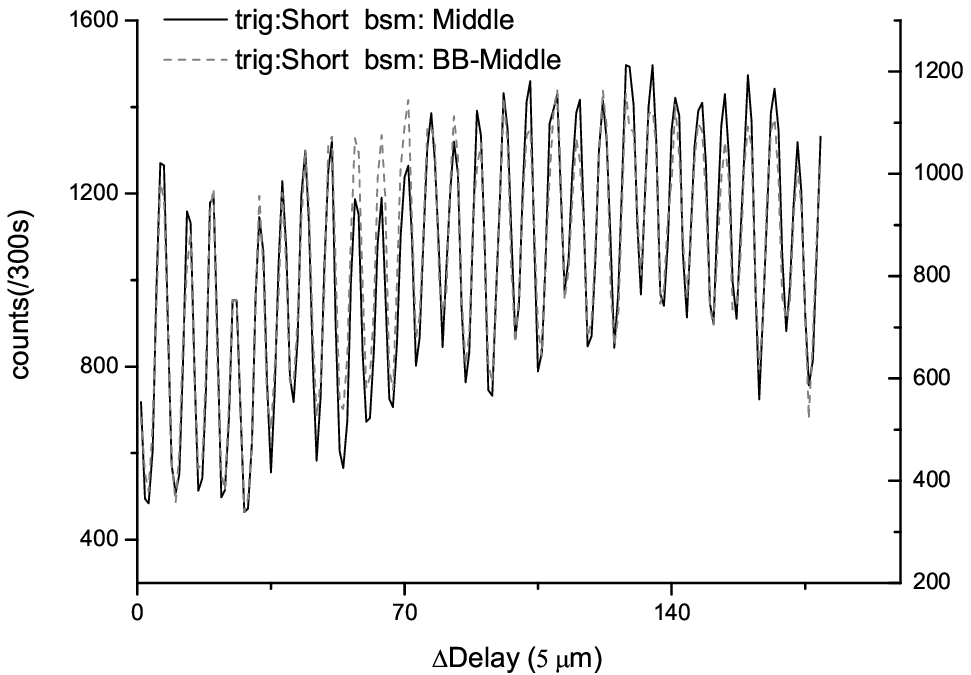} &
\includegraphics[width=4cm]{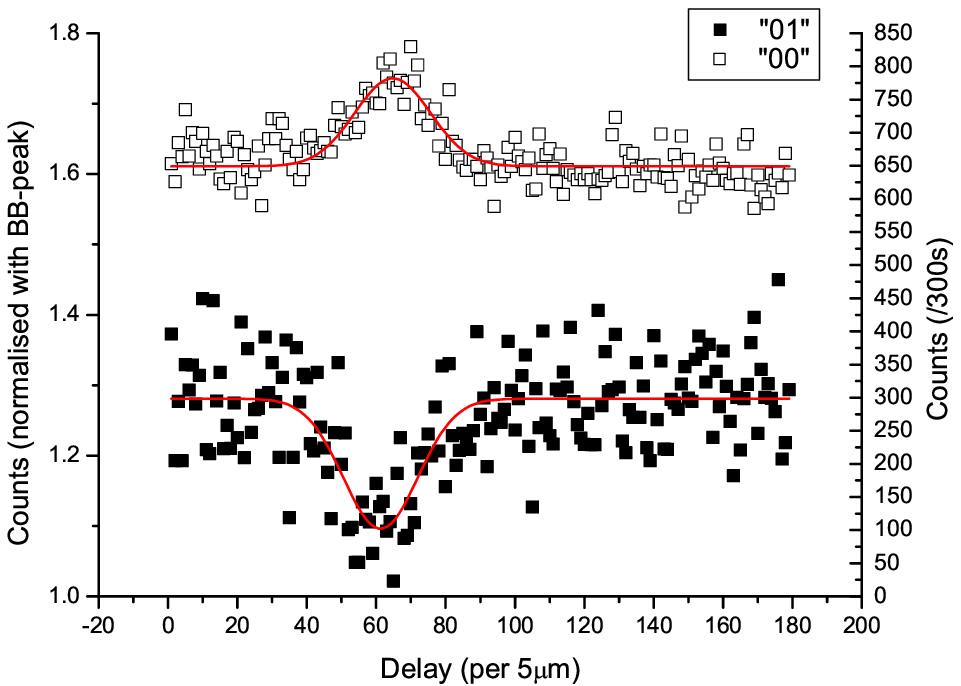}
\end{tabular}
\caption{\label{fig:dip2}\emph{Left:} The count rate for
\quote{01} depends on the phases of the interferometers, which
were not stabilized. Since this interference is single-photon
interference it is present for both the main
\quote{01}-coincidence and for the baby-peaks. \emph{Right:} Using
the baby-peaks to normalize the main count rate it is possible to
see a dip in the count rate when scanning an optical delay. This
decrease is caused by photon bunching. The dip is in the same
position as the measured antidip \quote{00}.}
\end{figure}

\p The antidips discussed above are not the only method of
aligning the setup. It is also possible to look at a dip. For
example there will be a decrease in the number of \quote{01}
depending on whether there is photon-bunching or not. During
measurements of such a decrease the interferometers are not
stabilized for experimental reasons. Since the coincidence-rate is
dependant on single-photon interference it is very difficult to
clearly see the decrease in counts \fig{fig:dip2}. One way to
avoid this problem is to use a baby-peak as a normalization.
baby-peaks are coincidences with one (or more) laser pulses of
difference between the creation time of the detected photons. For
example, laser-pulse $n$ creates a photon in Alice and this
photons goes to detector $D0$, while laser-pulse $n+1$ creates a
photon in the EPR-source which goes to $D1$. The amount of
coincidences measured for these pulses will depend on the
single-photon interference but there will clearly not be any
photon-bunching. Since such coincidances have the same
interference effects as for the real coincidences it can be used
to normalize a measurement and in this way a dip can be found
\fig{fig:dip2}. Since this normalisation method is much more
complicated and less accurate it was not used for alignment, only
antidip-alignment was used.

If temporal alignment isn't accomplished in a BS-BSA teleportation
experiment the resulting coincidence rates will not depend on the
phases of the interferometers and therefore a fringe with $V=0\%$
is found. When using an IF-BSA this is not the case since the
presence of the extra interferometer leads to single photon
interference when changing the phase $\alpha$. It is clearly
important to be able to distinguish between these interferences
and the interference fringes caused by teleportation. The behavior
of the non-aligned setup can be readily calculated and the fringes
that will be found are shown in Table \ref{table:distinguishable}.
One important fact clearly stands out straight away: there is no
interference for \quote{02} and \quote{20} if the photons are
distinguishable but there is when the photons are
indistinguishable. The visibility of these fringes are an
important indication whether or not there was temporal alignment
during the experiment. In the experiment presented here a
visibility of $V=55\% \pm 3\%$ was found which indicates that
there was temporal indistinguishability.

\begin{figure}[]
\includegraphics[width=8cm]{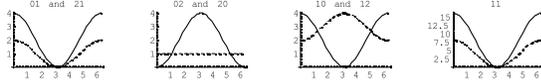}
\caption{\label{fig:simulation}Simulation of result from a
teleportation experiment in the case that the interferometers have
been aligned to have $\delta=-\beta$ as was the case in our
experiment. The dashed curves are for the case of distinguishable
photons at the BSA and the plain curves are for indistinguishable
photons (teleportation).}
\end{figure}

Other indications whether there is good temporal alignment can be
found when simulating the result of an unaligned experiment. Such
a simulation is shown \fig{fig:simulation} for the case of
$\delta=-\beta$ as was used during our experiments. The simulation
clearly shows differences between the two cases which are readily
identifiable in an experiment, such as the phaseshift of $\pi$
between the fringes for \quote{01} and \quote{10}. These
differences make it possible to see after an experiment whether or
not the alignment was good and stayed good.

\begin{table}
\begin{tabular}{l l l}
BSA & indistinguishable & distinguishable\\ \hline
\quote{01}& $1+cos(\alpha+\beta)$&$1+cos(\alpha)$\\
\quote{02}& $1-cos(\alpha+\beta)$&$1$\\
\quote{10}& $1+cos(\alpha+\beta)$&$3-cos(\alpha)$\\
\quote{11}& $4(1+cos(\alpha+\beta-2\delta))$&$2(2+2 cos(\alpha-\delta))$\\
\quote{12}& $1+cos(\alpha+\beta)$&$3-cos(\alpha)$\\
\quote{20}& $1-cos(\alpha+\beta)$&$1$\\
\quote{21}& $1+cos(\alpha+\beta)$&$1+cos(\alpha)$\\
\end{tabular}
\caption{\label{table:distinguishable}Theoretical interference for
different projections made by the BSA. Two different cases are
shown: indistinguishable photons (teleportation) and
distinguishable photons (noise). }
\end{table}

\bibliographystyle{prsty}
\bibliography{M:/BIB/GAP}

\end{document}